\documentclass{elsart}

\usepackage{epsfig}

\begin{document}

\begin{frontmatter}



\title{A realistic technique for selection of angular momenta from hot nuclei: A
case study with $^{4}$He  + $^{115}$In $\rightarrow$ $^{119}$Sb$^*$ at E$_{Lab}$ = 35 MeV}


\author[label1]{Deepak Pandit},
\author[label1]{S. Mukhopadhyay},
\author[label2]{Srijit Bhattacharya},
\author[label1]{Surajit Pal},
\author[label3]{A. De},
\author[label1]{S. R. Banerjee\corauthref{cor}}
\corauth[cor]{Corresponding author.}
\ead{srb@vecc.gov.in}


\address[label1]{Variable Energy Cyclotron Centre, 1/AF-Bidhannagar, Kolkata-700064, India}
\address[label2]{Department of Physics, Darjeeling Government College, Darjeeling-734101, India}
\address[label3]{Department of Physics, Raniganj Girls' College, Raniganj - 713358, India}

\begin{abstract}
A rather new approach employing Monte Carlo GEANT simulation for 
converting the experimentally measured fold distribution to angular 
momentum distribution has been described. The technique has been 
successfully utilized to measure the angular momentum of the compound 
nucleus formed in the reaction $^{4}$He  + $^{115}$In $\rightarrow$ $^{119}$Sb$^*$ 
at E$_{Lab}$ = 35 MeV. 
A 50 element gamma multiplicity filter, fabricated in-house, was used 
to measure experimentally the required fold distribution. The present method has 
been compared with the other ones exiting in the literature and relative 
merits have been discussed. 

\end{abstract}

\begin{keyword}
Multiplicity filter, BaF$_2$ scintillator, GEANT3 simulation
\PACS 24.10.Lx; 29.30.Kv; 29.40.Mc 
\end{keyword}
\end{frontmatter}

\section{Introduction}

The emission of $\gamma$-rays from the decay of giant dipole resonance (GDR) 
in hot and fast rotating nuclei provides a unique tool to study the various 
kinds of structure (triaxial, prolate, oblate, spherical) that the nuclear 
system can assume at high temperature (T) and angular momentum J \cite{Hara01, Gaard01, Snov01}. 
In heavy-ion fusion reaction, the compound nucleus is formed at well 
defined excitation energy, but with a wide range of angular momenta.  
The hot compound nucleus loses most of the excitation energy via particle 
and gamma emissions above the yrast line. The remainder of the excitation 
energy and angular momentum is generally removed by the low energy yrast 
gamma emission \cite{Hara01}. The GDR parameters depend on both excitation energy and 
angular momentum and to understand the individual contribution of T and J, 
it is important to separate the two effects. However, decoupling these two 
effects is a very difficult experimental task, the procedure adopted being the
measurement of the high energy photon spectrum in coincidence with the low energy 
gamma multiplicity. A precise measurement of this $\gamma$-multiplicity is 
very important since the number of $\gamma$-rays emitted is directly related 
to the angular momentum populated in the system. As per the usual techniques,
the multiplicity gamma rays 
are measured with an array of many detectors placed closed to the target 
having high efficiency and granularity. The fold (number of multiplicity 
detectors fired) distribution is recorded on an event-by-event basis in 
coincidence with the high energy gamma rays. Finally, the angular momentum 
distribution is extracted from this fold distribution in offline analysis. 
However, there is no straightforward procedure for mapping the fold 
distribution to angular momentum distribution and quite a few methods 
have been adopted in literature for converting the 
folds to J distributions  \cite{Mat97, Maj94, Drc06, Jas83}. 

In this paper, we present a rather new technique based on 
Monte Carlo GEANT3 \cite{geant} simulation for converting the 
fold distribution to angular momentum distribution. The approach 
has been tested for the reaction 
$^{4}$He  + $^{115}$In $\rightarrow$ $^{119}$Sb$^*$ at E$_{Lab}$ = 35 MeV, 
where the experimental fold distribution was measured with our recently 
fabricated gamma multiplicity filter. The above method has also been
compared with other approaches adopted earlier.

\section{The multiplicity spectrometer}

Recently, a 50-element gamma-multiplicity filter made of BaF$_{2}$ 
has been designed and developed at the 
Variable Energy Cyclotron Centre, 
Kolkata. 
The square shaped crystals have a cross-section of 
3.5$\times$3.5 cm$^2$ and 5 cm in length.
Standard procedures were followed for the fabrication of the
detector from bare barium fluoride crystals \cite{Supm}. 
Crystals were cleaned properly and then wrapped with 
several layers of white teflon tape, aluminium foil 
and black electrical tape.
Fast, UV sensitive photomultiplier tubes (29mm dia, Phillips XP2978) were 
coupled with the crystals using a highly viscous UV transmitting optical 
grease (Basylone, $\eta$ $\approx$ 300000 cstokes). 
Aluminium collars of unique shape were used around the 
coupling area to provide additional support. 
Finally, for mechanical stability, the whole assembly 
(crystal + PMT) was wrapped with heat shrinkable PVC tube.

\begin{figure}
\begin{center}
\includegraphics[height=9 cm, width=7.5 cm]{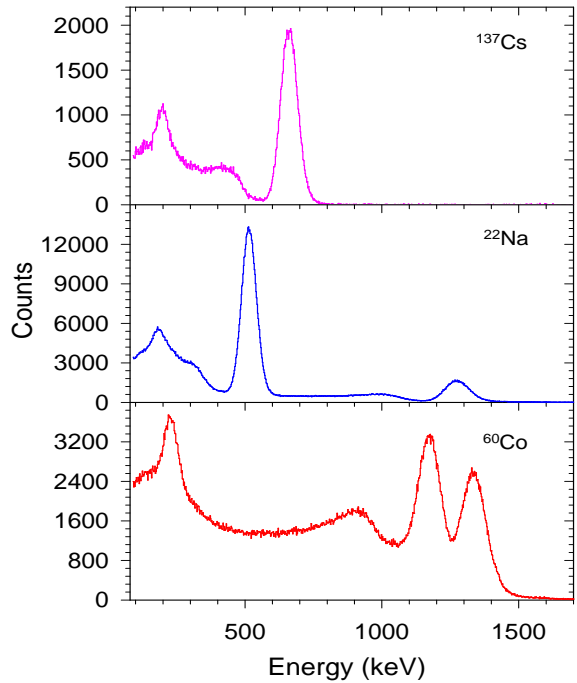}
\caption{\label{energy}Energy spectra from a single detector for different lab. 
standard gamma ray sources.}
\end{center}
\end{figure}

\begin{figure}
\begin{center}
\vspace{0.3cm}
\includegraphics[height=5.7 cm, width=7.5 cm]{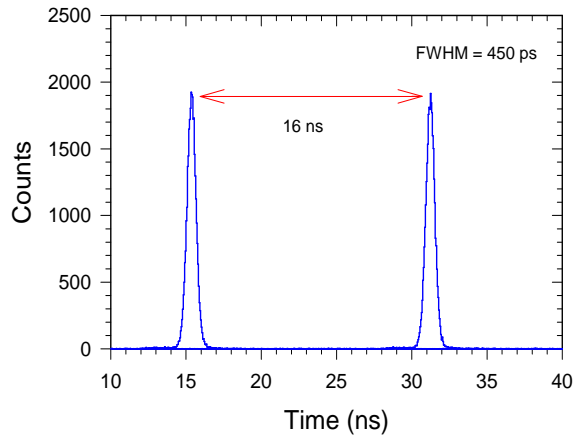}
\caption{\label{time}Time resolution of an individual detector using $^{60}$Co $\gamma$-source.}
\end{center}
\end{figure}

After fabrication, the individual detector elements were tested using 
standard gamma ray sources. Typical experimental energy spectra for an individual 
detector is shown in Fig. \ref{energy}.
The observed energy resolution is 7.2$\%$ at 1.17 MeV. 
The time resolution between two BaF$_{2}$ detectors was measured with the $^{60}$Co source. 
The source was placed in between two identical detectors, which were kept 180$^{\circ}$ apart. 
The energies and their relative times were measured simultaneously in event by event mode. 
The resulting energy gated (1.0-1.4 MeV) time spectrum is shown in Fig. \ref{time}. 
The value obtained for time resolution is 450 ps.

\section{In - beam experiment }

\begin{figure}
\begin{center}
\includegraphics[height=6.0 cm, width=7.5 cm]{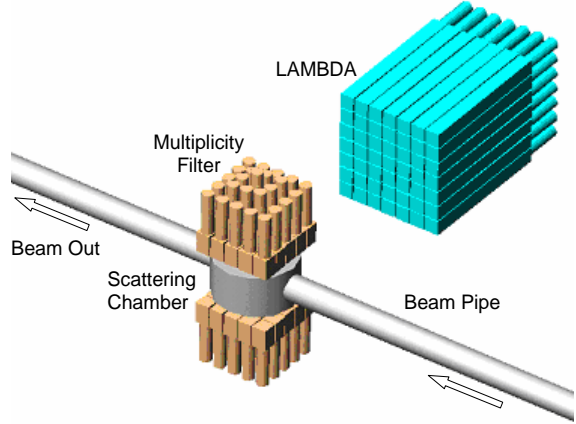}
\caption{\label{setup} Schematic view of the experimental setup.}
\end{center}
\end{figure}

The in-beam performance of the multiplicity filter was tested 
using alpha beam from the K-130 AVF cyclotron at VECC.
A 1 mg/cm$^{2}$ target of $^{115}$In was bombarded with 35 MeV alpha 
beam producing $^{119}$Sb at 36 MeV excitation energy (L$_{cr}$=16$\hbar$). 
For the estimation of the angular momentum populated by the 
compound nucleus, the 50-element filter was split into two blocks of 25 detectors 
each and was placed on the top and the bottom of the 
scattering chamber at a distance of 5 cm from the 
target center (covering 56$\%$ of 4$\pi$) in castle geometry. 
The detectors of the multiplicity filter were gain matched
and equal threshold was applied to all. 
Along with the filter, a part of the LAMBDA 
spectrometer \cite{Supm} (49 large BaF$_2$ detectors arranged in 7$\times$7) was also used to 
measure the high energy gamma rays ($>$ 4 MeV) in coincidence 
with low energy discrete gamma rays. The high energy photon 
spectrometer was centered at 90$^\circ$ to the beam direction and 
at a distance of 50 cm from the target. The schematic view 
of the experimental setup is shown in Fig. \ref{setup}. 
The detectors of the multiplicity filter in castle geometry 
were staggered in order to have equal solid angle for each 
detector in the array.

\begin{figure}
\begin{center}
\includegraphics[height=5.5 cm, width=7.5 cm]{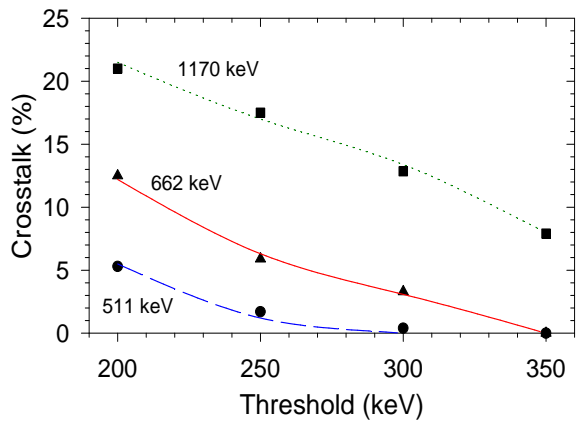}
\caption{\label{xtalk} The crosstalk probabilities for three energies at different 
thresholds of the multiplicity filter. The symbols represent the experimental points 
while the lines correspond to GEANT simulation.}
\end{center}
\end{figure}

A level-1 trigger (A) was generated from the multiplicity filter array when 
any detector of the top block and any detector from the bottom block
fired in coincidence above a threshold of 250 keV. Another trigger (B) was generated 
when the signal in any of the detector elements of the LAMBDA 
spectrometer crossed a high threshold ($>$ 4 MeV). A coincidence 
of these two triggers generated the master trigger ensuring 
the selection of the high energy photon events from the 
compound nucleus and rejection of background. The sum of 
the multiplicity filter (number of detectors fired in the event) 
was fed into a QDC (V792) gated 
by the master trigger to generate, on event-by-event basis, 
the experimental fold(F) distribution with condition (F $\geq$ 2). 
The crosstalk probability of the multiplicity set-up was also measured
using $^{22}$Na (511, 1274 keV), $^{137}$Cs (662 keV) and $^{60}$Co (1.17, 1.33 MeV) 
sources at different discriminator thresholds of the filter.
The measured crosstalk probabilities are shown in Fig. \ref{xtalk}(symbols). 
It is observed that the scattering probability is more at higher 
energies and it decreases with increasing the threshold of the multiplicity detectors.  

\section{Existing procedures for determining the angular momentum}

In general, the multiplicity distribution method is widely used 
to convert the experimentally measured fold distribution into angular 
momentum distribution \cite{Mat97, Maj94}. In this method, the low energy gamma 
multiplicity (M) is derived from the measured fold (F) distribution where the 
M to F response function of the multiplicity array is measured 
experimentally. The experimental procedure consists of placing a 
source at the target center emitting 2 gamma rays in  cascade (e.g $^{60}$Co) 
and recording the gamma rays in the multiplicity filter. An external 
detector is used as a trigger, and the events are collected 
by selecting the photo-peak of 1.33 MeV gamma rays in it from the $^{60}$Co source,
ensuring that exactly one gamma ray (1.17 MeV) is incident to the filter. 
With this condition, the events consisting of the analog signal 
proportional to the fold are stored in the list mode. 
Hence, the collected fold spectrum is the response of the 
filter to the $\gamma$-ray multiplicity M=1 (at specific energy 1.17 MeV). 
The response to the multiplicity M=k is generated, in offline analysis, 
by randomly selecting k events from the previously stored data in list 
mode and summing up the amplitudes of the associated individual fold signals. 
The M distribution is assumed to have a gaussian or a triangular form 
and the corresponding F-distribution is calculated by folding it with the 
above response function. The parameters of the M-distribution are varied 
until the best fit to the measured F-distribution is obtained. 
After getting the full M-distribution, the constraint M-distributions 
are calculated for the different F-windows.

\begin{figure}
\begin{center}
\includegraphics[height=8 cm, width=7.5 cm]{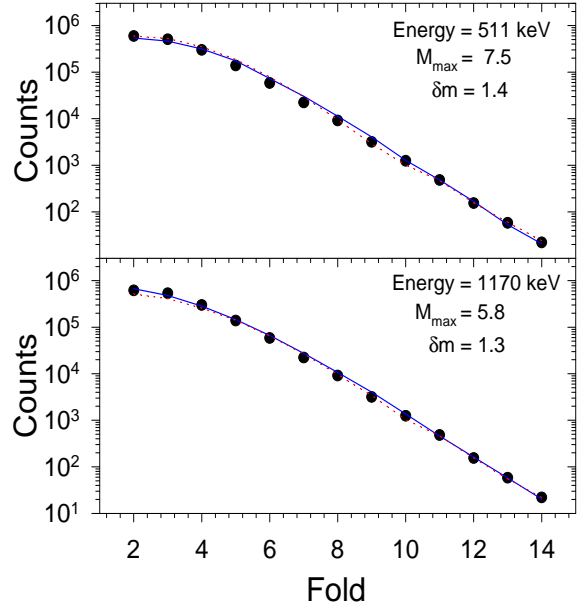}
\caption{\label{f511} The filled circles represent the experimentally measured fold distribution.
The solid line represents the fold distribution obtained from the multiplicity distribution method
while the dotted line corresponds to GEANT simulation. }
\end{center}
\end{figure}

\begin{figure}
\begin{center}
\includegraphics[height=4.5 cm, width=7.5 cm]{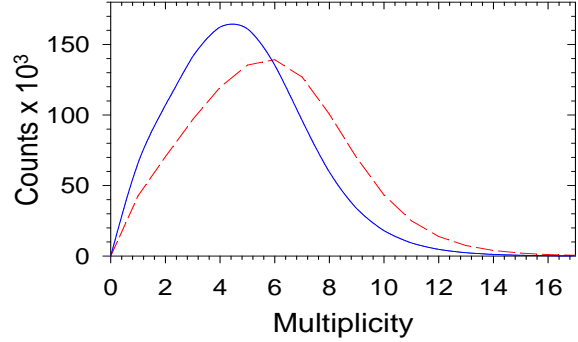}
\caption{\label{multi} The continous line represents the triangular distribution obtained for 1.17 MeV response funtion
while the dashed line corresponds to the 511 keV response function. }
\end{center}
\end{figure}

In order to test the reliability of the method, the experimentally 
measured fold was converted to the multiplicity M using the above formalism.
To remove the contribution of non-fusion events,
the final experimental fold spectrum was generated, offline, by gating with 
high energy gamma rays ($>$ 10 MeV) \cite{Sri}.
Following the multiplicity distribution method, the response function of the multiplicity
filter was created for 1.17 MeV using $^{60}$Co. 
Similarly, the response function for 511 keV was created using $^{22}$Na source.
The multiplicity distribution was assumed triangular as follows:- 

\begin{equation}\label{mul}
P(M)=\frac{2M+1}{1+exp[(M-M_{max})/\delta m]}
\end{equation}
where, M$_{max}$ is the maximum of this distribution and $\delta$m is the diffuseness.
The multiplicity distribution was folded with the response function to generate the corresponding
fold distribution. The parameters M$_{max}$ and $\delta$m were varied in order to match 
the experimental fold distribution. The comparison between the experimental fold distribution
and those obtained using the multiplicity distribution method is shown in Fig. \ref{f511}.
Interestingly, the M$_{max}$ and $\delta$m values 
extracted using the two-response functions (for two different energies, 511 keV and 1.17 MeV) are quite different.
For 511 keV the values of M$_{max}$ and $\delta$m are 7.5 and 1.4 respectively, whereas 
for 1.17 MeV, the corresponding values are 5.8 and 1.3. The difference between the two triangular distributions
is clearly seen in Fig \ref{multi}. This difference is due to the fact that the scattering probability and 
efficiency of the filter for the two energies is different (Fig \ref{xtalk}). 
Consequently, the constraint M-distributions for different F-windows will be
different for the two response functions. Moreover, the energy of the multiplicity 
gamma rays are not constant and depend on the initial and final J of a given transition. 
Therefore, generating the response function of the multiplicity filter at single energy 
will give incorrect values of average J for corresponding F windows as both, efficiency 
and scattering probability, depends on the gamma energy.
Ideally, the energy distribution of the emitted multiplicity gamma rays should be 
measured experimentally and the response function should be created according to the measured energy distribution. 
However, calibrating the filter with different energy is experimentally very difficult 
as the sources emitting two gamma rays in cascade are not available always.
Moreover, selecting k events from different response functions according 
to the energy distribution of the $\gamma$ multiplicity will also be a very complicated job. 
As a result, for generating a realistic 
response function of the multiplicity filter to incorporate the energy dependence of 
efficiency and scattering probability, 
the only possible procedure is a Monte Carlo simulation.

Another approach, the recursion method \cite{Maj94}, has also been adopted 
in literature to convert the fold to multiplicity distribution. 
In this method, the probability P(F,M) of triggering F out of N detectors by a 
cascade of M $\gamma$-rays can be calculated by using a simple recursive 
algorithm. The input parameters of the recursion are the total efficiency 
and scattering probability. This method gives practically identical results 
as obtained from the multiplicity distribution method  \cite{Maj94}. 
The above formalism also suffers from the same problem since the efficiency 
and the scattering probability depends on the $\gamma$ energy.

\begin{figure}
\begin{center}
\includegraphics[height=5 cm, width=7.5 cm]{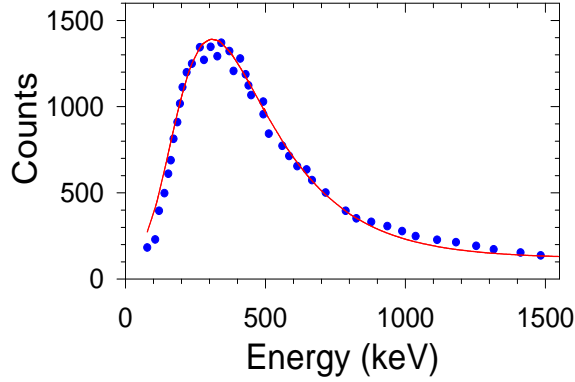}
\caption{\label{in_ener} The experimentally measured  energy distribution
of the $\gamma$-multiplicity (symbol) fitted with Landau function as used in GEANT simulation.}
\end{center}
\end{figure}

\section{The present approach for determining the angular momentum}

In this section, we describe an approach based on Monte Carlo 
GEANT3 \cite{geant} simulation for conversion of the experimental fold distribution to the 
angular momentum distribution.
In this simulation, the realistic experimental conditions 
(including the detector threshold and trigger condition)
are taken into account. The consistency of our simulation was checked 
by generating the fold distribution for the energies of 
1.17 MeV and 511 keV considering the same parameters for incident 
gamma multiplicity as used earlier for multiplicity distribution method.  
Two blocks of 25 detectors arranged in 5 $\times$ 5 arrays  
were kept on the top and bottom of the scattering chamber, similar to the experiment, 
at a distance of 5 cm from the target position.
The different input multiplicities of the low energy $\gamma$ - rays 
were obtained by creating a random number according to the multiplicity 
distribution P(M). Low energy gamma rays, for each randomly generated 
multiplicity, were thrown isotropically from the target centre and the 
corresponding fold was recorded for that event. Two hundred thousand 
such events were triggered to record the final simulated fold distribution
considering single energy of the incident gamma rays (511 keV and 1.17 MeV).
The fold distribution obtained using 
GEANT simulation and those obtained from the multiplicity distribution method match 
quite well with each other (dotted line in Fig \ref{f511}). 
Earlier, the simulated scattering probability was also found to be in good
agreement with the experimental observation (Fig \ref{xtalk}).

\begin{figure}
\begin{center}
\vspace{0.8cm}
\includegraphics[height=5 cm, width=7.5 cm]{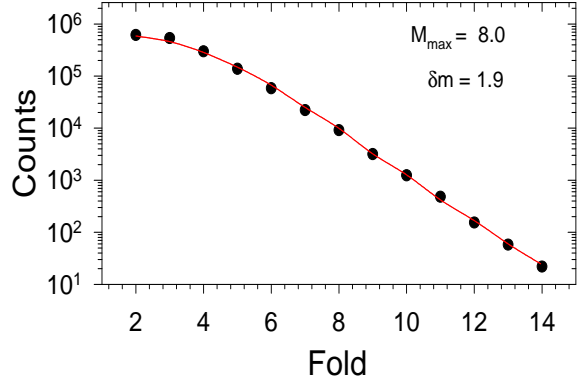}
\caption{\label{fgeant} Experimental fold spectrum (symbols) fitted with GEANT simulation (solid line).}
\end{center}
\end{figure}

\begin{figure}
\begin{center}
\includegraphics[height=5 cm, width=7.5 cm]{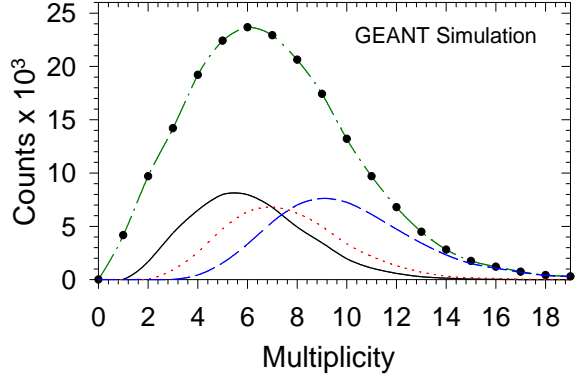}
\caption{\label{gmult} The incident multiplicity distribution used in GEANT simulation (symbols along with dot-dashed line). 
The multplicity distributions obtained for different folds are also shown in the figure.
The solid line represents fold 2, the dotted line represents fold 3 and the dashed line,
the multiplicity distribution for folds 4 and more.}
\end{center}
\end{figure}

\begin{figure}
\begin{center}
\includegraphics[height=5 cm, width=7.5 cm]{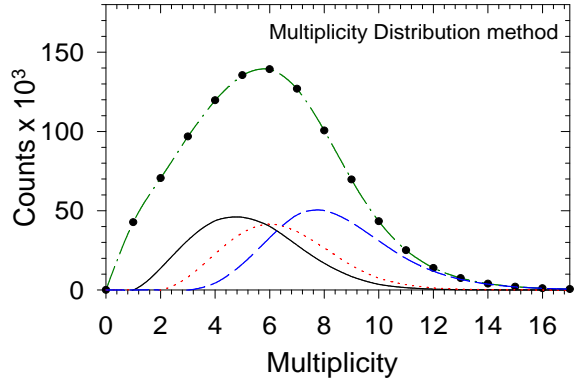}
\caption{\label{cmult} The incident multiplicity distribution used in multiplicity distribution method
using 511 keV response function(symbols along with dot-dashed line). The solid line represents fold 2, the dotted line represents fold 3 and the dashed line, the multiplicity distribution for folds 4 and more.}
\end{center}
\end{figure}

In order to have the correct energy distribution  
for simulation, the energies of the gamma multiplicity were measured 
experimentally. The distribution is shown in Fig. \ref{in_ener} (filled circles). 
The angular momentum distribution for this reaction was obtained from the statistical model code CASCADE \cite{cas}.
The conversion of the angular momentum distribution to multiplicity distribution is 
achieved using the relation J = 2M + C, where C is the free parameter which takes 
into account the angular momentum loss due to particle evaporation and emission of 
statistical $\gamma$-rays. The final simulated fold distribution was generated using the 
multiplicity distribution along with its measured energy distribution.
The incident energy distribution was  parameterized by a 
Landau function (continous line in Fig. \ref{in_ener}) given as 

\begin{equation}\label{lan}
L(E_\gamma)=n\sqrt{\frac{e^{-(p+e^{-p})}}{2\pi}}
\end{equation} 
where p=c$\cdot$(E$_{\gamma}$-b), n=5350, b=320 keV and c=0.0085. 
The parameters M$_{max}$ and $\delta$m of the
multiplicity distribution was obtained from the J-distribution by varying the 
free parameter C until the best fit to 
the measured F-distribution was
achieved. The value of C was obtained as 0.5 and the parameters of the 
M-distribution were extracted as M$_{max}$ = 8.0  and $\delta$m = 1.9 for best fit.
The extracted value of C seems to be reasonable as the angular momentum loss due to 
particle emission will be negligible (since medium mass nuclei is 
populated at low excitation energy). Also, the experimental fold distribution 
was generated by gating with high energy gamma rays ($>$ 10 MeV), which further reduces 
the average angular momentum loss. 
The simulated fold distribution generated using the above triangular 
distribution is shown by solid line in Fig. \ref{fgeant}.
Next, the constraint multiplicity distributions for different folds were generated. 
The incident multiplicity distribution (dot-dashed line along with symbol)  and the multiplicity distributions 
for different fold windows are shown in Fig. \ref{gmult}. The continuous line, 
the dotted and the dashed lines indicate the multiplicity distributions gating 
on the events with folds 2, folds 3 and folds $\geq$ 4 respectively. 
The above results have been compared with the multiplicity distribution 
method for 511 keV and 1.17 MeV response function. The incident multiplicity and 
the multiplicity for different folds (511 keV response function) are shown in Fig \ref{cmult}. 
The average angular momentum values for 
different fold windows are summarized in table 1.

As could be discerned, the results obtained from the present method with respect to multiplicity distribution method
(using 511 keV response function) differs by $\sim$ 15$\%$. This is also expected because the average energy of the multiplicity $\gamma$-rays is less than the single energy (511 keV) used in the multiplicity distribution method. Thus, it seems to be important that the energy dependence of the efficiency and the cross talk probabilities of the filter should be taken into consideration while converting the measured fold distribution into corresponding angular momentum distribution.

\begin{table}

\caption[]{\label{tab:nucl} Average angular momentum values corresponding
to different folds as obtained from GEANT simulation (present work) and from multiplicity distribution method
calibrated at two different energies (511 keV \& 1.17 MeV).}
\begin{center}		
\begin{tabular}{|c|c|c|c|}
\hline
Fold     & GEANT Simulation & Multiplicity Distribution & Multiplicity Distribution \\
 &       &           (511 keV)   &    (1.17 MeV) \\
 &	$\left\langle J \right\rangle$$\hbar$  & $\left\langle J \right\rangle$$\hbar$ & $\left\langle J \right\rangle$$\hbar$ \\
\hline
2 &     12.2 $\pm$ 4.7   &    10.5 $\pm$ 4.5  &    9.2 $\pm$ 4.3   \\
\hline
3 &     15.1 $\pm$ 4.8    &  12.9 $\pm$ 4.5  &  11.0 $\pm$ 4.3 	  \\
\hline
 4 $\&$ more  &  20.1 $\pm$ 5.2    &  16.8 $\pm$ 4.9  &  14.1 $\pm$ 4.7  	\\
\hline
\end{tabular}
\end{center}		
\end{table}

\section{Summary}

An approach based on Monte Carlo GEANT3 simulation has been presented 
for the selection of angular momentum space from the experimentally measured 
fold distribution. Drawback inherent in the existing approaches has been discussed 
and the present method has been applied to overcome the same. Experimental fold 
distribution was obtained employing our recently fabricated 50-element BaF$_2$ 
multiplicity filter in the reaction $^{4}$He  + $^{115}$In $\rightarrow$ $^{119}$Sb$^*$  at 35 MeV beam energy. 
The present approach seems to have a significant importance while selecting 
the angular momentum space properly.



\end{document}